\documentclass[a4paper]{spie}  %>>> use this instead for A4 paper
\usepackage[]{graphicx}
\usepackage{amstext} 
\usepackage{amssymb} 
\usepackage{amsmath}
\usepackage{txfonts}
\usepackage{hyperref} 
\usepackage{url} 

\usepackage{caption}
\usepackage{subcaption}

\newcommand*\aap{A\&A}

\newcommand*\aj{AJ}

\newcommand*\apjl{ApJ}

\newcommand*\apss{Ap\&SS}

\newcommand*\nat{Nature}

\title{Enabling Gaia observations of naked-eye stars}
\author{J.~Mart\'in-Fleitas\supit{ab}, A.~Mora\supit{ab}, J.~Sahlmann\supit{a},  R.~Kohley\supit{a}, B.~Massart\supit{c}, J.~L'hermitte\supit{c}, M.~Le Roy\supit{c}, P.~Paulet\supit{c} 
\skiplinehalf
\supit{a}European Space Agency, European Space Astronomy Centre, P.O. Box 78, Villanueva de la Ca\~nada, 28691 Madrid, Spain;\\
\supit{b}Aurora Technology, Crown Business Centre, Heereweg 345, 2161 CA Lisse, The Netherlands;\\
\supit{c}Airbus Defence and Space, 31 Avenue des Cosmonautes, Z.I. du PALAYS, 31402 Toulouse Cedex 4, France}
\authorinfo{Further author information: (Send correspondence to J.M.F.)\\J.M.F.: E-mail: jfleitas@sciops.esa.int}%%%%%%%%%%%%%%%%%%%%%%%%%%%%%%%%%%%%%%%%%%%%%%%%%%%%%%%%%%%%% 
\begin{document} 
  \maketitle 

%%%%%%%%%%%%%%%%%%%%%%%%%%%%%%%%%%%%%%%%%%%%%%%%%%%%%%%%%%%%% 
%%%%%%%%%%%%%%%%%%%%%%%%%%%%%%%%%%%%%%%%%%%%%%%%%%%%%%%%%%%%% 
\begin{abstract}
The ESA Gaia space astrometry mission will perform an all-sky survey of stellar objects complete in the nominal magnitude range $G$ = [6.0 - 20.0]. The stars with $G$ $<$ 6.0, i.e. those visible to the unaided human eye, would thus not be observed by Gaia. We present an algorithm configuration for the Gaia on-board autonomous object observation system that makes it possible to observe very bright stars with $G$ = [2.0-6.0). Its performance has been tested during the in-orbit commissioning phase achieving an observation completeness of $\sim 94\%$ at $G=3$~--~5.7 and $\sim 75\%$ at $G=2$~--~3. Furthermore, two targeted observation techniques for data acquisition of stars brighter than $G$=2.0 were tested. The capabilities of these two techniques and the results of the in-flight tests are presented. Although the astrometric performance for stars with $G$ $<$ 6.0 has yet to be established, it is clear that several science cases will benefit from the results of the work presented here.

\end{abstract}

%%%%%%%%%%%%%%%%%%%%%%%%%%%%%%%%%%%%%%%%%%%%%%%%%%%%%%%%%%%%%
\section{INTRODUCTION}\label{sec:intro}
Gaia\cite{Perryman:2001vn} , the astrometry mission of the European Space Agency, was launched on 19$^{th}$ December 2013 and its commissioning phase ended in mid-2014\cite{2014SPIE.Prusti,2014SPIE.Els} . Gaia is a spinning spacecraft designed to observe every star $\sim$70 times during the 5-year mission lifetime in the magnitude range $G$ = [6.0 - 20.0]\footnote{The relationship between $G$ and $V$ can be found in [\citenum{2010A&A...523A..48J}].}, being partially complete only in very high-density areas of the sky. Gaia will create the most detailed 3D map of the Galaxy, comprising a billion stars, providing astrometric, photometric, and spectroscopic data. Gaia improves every single aspect of its predecessor, the Hipparcos space mission\cite{Perryman:1989kx} , except for the bright star magnitude limit.

The Gaia spacecraft observes with two wide-field telescopes sharing a common focal plane. Gaia's Focal Plane Array (FPA) is the largest flown in space so far, with 106 CCDs organised in 7 rows and 17 strips, a total of almost 1,000 million pixels, and physical dimensions of 1.0 m $\times$ 0.45 m.

Each of the 7 CCD rows is controlled by one Video Processing Unit (VPU). The VPUs are hybrid microprocessor-FPGA systems in charge of autonomously detecting, confirming, and observing any star-like object crossing the focal plane in the nominal magnitude range. The bright-end magnitude limit is defined by the VPU parameter configuration. By optimising this configuration, we show that the bright limit of on-board source confirmations can be improved, although some extremely bright stars remain beyond the system's capabilities. To allow data gathering of these stars as well, two targeted observation techniques were designed and tested. These techniques are the SIF imaging, which allows taking images of the stars, and the VO synchronised observations, which allows taking a 'nominal' observation of the object, i.e. imaging, photometry, and spectroscopy data.

We discuss the scientific motivation for observing the naked-eye stars (Sect.~\ref{sec:sci}), introduce the Gaia detection system for the particular case of bright stars (Sect.~\ref{sec:obs}), and present the VPU algorithm configuration proposed (Sect.~\ref{sec:vpu}). The two targeted observation techniques, SIF imaging (Sect.~\ref{sec:sif}) and VO synchronised observations (Sect.~\ref{sec:vo}), are presented and the results from the commissioning tests are reported. We conclude with an outlook to the results and to the forthcoming work (Sect.~\ref{sec:further} and Sect.~\ref{sec:conclusion}).

\section{SCIENCE Opportunity}\label{sec:sci}
The naked-eye stars played a fundamental role in the development of astronomy and astrophysics, because they can be studied with the unaided human eye, and detailed investigations are possible with small and intermediate-size telescopes. Precise distances to most of these stars were largely unknown prior to the Hipparcos mission in 1989-1992, whose parallax and proper motion measurements remain unique still today. Since naked-eye stars are the brightest, they often contain the most nearby example of a certain stellar type, thus benchmarks for the understanding of star formation and evolution. Several of them were resolved with stellar interferometry to measure stellar diameters and shapes, they serve as targets for photometric studies for asteroseismology and transiting exoplanet search, and they contain Vega and Sirius.

The Hipparcos catalogue contains about 4800 stars with magnitudes of $G<5.7$\footnote{To ensure completeness at $G=6$, Gaia's onboard detection algorithm was designed to be efficient for stars up to $G=5.7$. Here, we therefore concentrated on stars brighter than the latter limit.}. The astrometry of these stars has been monitored over $\sim3$ years around 1990 with an accuracy usually not better than 1 mas \cite{ESA:1997vn, :2007kx}. From the ground, it remains difficult to obtain astrometry at this level for bright stars \cite{Shao:1990qq}, in particular if one wants to target the complete sample. The only scheduled space mission for astrometry of bright stars is nano-Jasmine \cite{Yamada:2013aa}, which aims at $\sim$3 mas accuracy.

Obtaining the astrometry of stars with $G<5.7$ therefore represents a unique opportunity for Gaia, which is the motivation for the work presented here. These stars were initially not part of the nominal mission because of the requirement to be complete out to $G=20$ and dynamic range constraints on the system design. Yet, there are several science cases for Gaia astrometry of very bright stars\cite{LL:JSA-001}. The fundamental one is the improvement in the accuracy of their parallaxes and proper motions by at least one order of magnitude. The parallaxes are particularly valuable, because in many high-precision measurements of stellar parameters, the knowledge of the distance to a star is the limiting factor. 

Very bright stars are prime targets for extrasolar planet search with all available techniques. Many planets discovered through radial velocimetry suffer from the $\sin i $-ambiguity of their masses. Gaia astrometry can determine the inclination $i$ of the orbital plane and thus tightly constrain the planet mass. For instance, 47 UMa is a $G\simeq4.9$ star and hosts one of the first exoplanets to be discovered\cite{Butler:1996aa} with a minimum mass of $\sim$2.5 MJ. The minimum astrometric signature caused by this planet is $\sim$350 $\mu$as. If Gaia astrometry at the 100 $\mu$as level can be obtained for this star, its orbital motion due to planet b will be detected and the ambiguity in the planet mass will be resolved. The Gaia observations will also make is possible to search for new giant planets around very bright stars, with the advantage of being independent of the star's spectral type or age. Finally, the orbital motions of very bright binary stars can be measured with Gaia, leading to an improved knowledge of their physical parameters.

A detailed description of the science opportunities enabled by Gaia observations of naked-eye stars is being prepared\cite{Sahlmann:2014}.

\section{Gaia Observation system}\label{sec:obs}
The two field of views (FOV) of Gaia's telescopes are combined and observed in a single Focal Plane Array. The FPA\cite{2007SPIE.6690E...8L,2012SPIE.8442E..1PK} comprises 106 CCDs mounted on a Common Support Structure aligned in 7 rows and 17 functional strips, with a total size of $~$1m $\times$ $~$0.45 m. Each CCD has 4500 pix $\times$ 1966 pix, with a physical size of 10~$\mu$m ~$\times$ ~30 ~$\mu$m and an equivalent angular size on the sky of 59 mas $\times$ 177 mas per pixel. Each CCD is operated in TDI mode synchronized with the spacecraft's 6 hour spinning period.

The FPA is divided in 4 main instruments\cite{de-Bruijne:2012kx}. The Sky Mapper (SM), comprising 14 AF-Type CCDs (seven per FOV) continuously read to detect the objects to be observed. The Astrometric Field (AF) contains 62 AF-Type CCDs distributed in nine strips. The AF measurements are devoted to providing the astrometric information for the observed stellar source. The photometer comprises the Blue Photometer (BP, 7 BP-Type CCDs) covering the 330--680 nm wavelength range with a resolution of R=20--70 and the Red Photometer (RP, 7 RP-Type CCDs) covering the 640--1050 nm wavelength range with a resolution of R=60--90. Finally, the Radial-Velocity Spectrometer (RVS) contains 12 RP-Type CCDs and covers the 847--874 nm wavelength range with a resolution of R$\sim$11.500. 

Each of the 7 CCD rows of the FPA is controlled by a Video Processing Unit (VPU). The VPUs\cite{LL:VPU.00001} are hybrid systems composed {of two boards,} a custom designed board with two Actel FPGAs and three SDRAM memory slots (the "Companion Board"). The hardware processing tasks are time critical and are executed on this board. And a processor board from  Maxwell Technologies (SCS750 Single Board Computer for Space) with three TMR-protected PowerPC 750 FX processors, error detection, re-synchronisation, processor scrubbing, and a maximum processing speed of 800 MHz. Its main memory comprises one SDRAM (256 MB) and one ROM (8 MB). This board takes care of all the other algorithmic processing, called the software processing.

The VPU is in charge of detecting, confirming, and observing objects in the nominal magnitude range that cross the focal plane in all instruments. To limit the amount of telemetry stored on-board and downloaded to ground, the VPU decides in real time which objects to observe and how to distribute the available resources. The behavior of each VPU can be controlled independently by changing the configuration of the VPU algorithms, which are highly parametrised.

Every object crossing the focal plane is first observed either by SM1 or SM2, coming from telescope 1 or telescope 2, respectively. The SMs are read in  {Full Readout Mode}, generating Full Video Packages (FVP) stored in the SM Raw Data Buffers (one buffer per SM). By analyzing these data, the VPU HW algorithms detects the objects, determines its centroid location, computes its on-board magnitude ($g\_mag$\footnote{$g\_mag$\cite{LL:GAIA.ASF.UM.PLM.00022} is the on-board magnitude estimate of an object. It is coded in 10 bits and its relationship with $G$ is: $G = 21 - (g\_mag/64)$.}), determines its correspondent class of sampling\footnote{The class of sampling defines the combination of window size and binning to be assigned to the observation. There are three classes of sampling named from 0 to 2 (the photometer has an additional class of sampling). Their exact definition depends on the FPA instrument the object is transiting\cite{LL:GAIA.ASF.UM.PLM.00022}. The very bright stars studied here are assigned class\_0 in all the instruments.} and, finally, assigns an observation window to the object. The window is then propagated through the following CCDs of the CCD row as the imaged object transits the eight next CCD strips in AF, then the BP, RP, and RVS (the latter are present only for four of the seven CCD rows). The actual propagation uses input from the spacecraft's attitude control system, from which the VPUs predict the position of each object in the focal plane as a function of time.

The VPU uses the data gathered on the first strip of the astrometric field (AF1) to confirm an object detection in SM. This confirmation process is mainly aimed at discarding observations of false SM-detections, e.g. caused by cosmic rays. From AF1 onwards, {the CCDs are read in Windowing Mode,} therefore, the VPU reads only the information inside the assigned windows, generating a Sample Video Package (SVP) for each CCD on the row and storing them in the SVP Raw Data Buffer. The windowed read-out configuration reduces both the readout noise to a few electrons and the amount of telemetry generated per object.

For every observed object, the VPU collects the data in Star Packets (SP) of two types: SP1 containing the SM, AF, BP and RP data and SP2 containing the RVS data. Once downlinked, the data of these SPs is processed by the downlink processing pipeline\cite{2014SPIE.Siddiqui} and finally distributed to the data processing centers and members of the Gaia Data Processing and Analysis Consortium (DPAC)\cite{LL:FM-030, 2008IAUS..248..224M, 2007ASPC..376...99O} .

\subsection{On-board bright star observation}\label{sec:BSobs}
The detection capability at the bright magnitude end is mainly determined by the saturated-star observation algorithms in the VPU and the configuration parameters that control its behaviour. The central samples of a bright star are saturated in the SM. The HW processing algorithms are not able to estimate the flux nor the centroid for stars with saturated samples {inside the 3$\times$3 samples of the inspection window used in SM (one SM-sample measures 2$\times$2 pixels because SM CCDs are operated with 2$\times$2 pixel on-chip binning). However, using a bigger inspection window of 5$\times$5 samples, the algorithm can detect the samples located close to border of the saturated area, those having saturated samples only on some of the external pixels of the big inspection window. Those are flagged as "extremity" and are stored in four separate buffers according to the relative location of the saturated samples with respect to the non-saturated ones}: North, South, East and West (N, S, E and W, respectively). The detection of a bright star is accomplished by the SM SW processing algorithms that cross-match those extremities, i.e. N with E and/or W, and S with E and/or W, and that determine which of them originate from the same bright star. The magnitude of the star is obtained from the magnitude computed for the corresponding extremities. A detailed description of the saturated star observation algorithms and the cross-matching process can be found in the VPU SW specification \cite{LL:GAIA.ASF.SP.PLM.00073}. The main parameters defining the behavior of the VPU saturated-star observation algorithms are:

\begin{enumerate}	
	\item The maximum number of extremities in any of the 4 extremity buffers is derived from the value of the VPU parameter  "EXT\_IN\_BUFFER". If the number of extremities in a buffer is equal to the maximum number of extremities allowed, no new extremity is added to that buffer.
	\item The life time of an extremity in any of the four buffers is derived from the value of VPU parameter "EXT\_WH", which effectively establishes the working horizon for the matching of extremities. Any extremity {laying} beyond the maximum life time is discarded.
	\item The magnitude of an extremity is computed from the flux measured in the SM and a correction that is based on another two parameters: "MAG\_AL\_LUT" and "MAG\_AC\_LUT". These LUTs (Look Up Table) give an estimate of the magnitude offset to add to the estimated extremity magnitude as a function of its distance to the centre of the saturated star (Along-Scan (AL) and Across-Scan (AC)) and the type of extremity. Implicitly, these LUTs depend on the assumed PSF definition.
	\item The maximum accepted magnitude difference between two extremities assigned to the same star is determined by the value of the VPU parameter "DELTA\_MAG".
\end{enumerate}

\begin{table}[h!]
\caption{VPU parameters configuration.}
\label{tab:Param}
\begin{center}
\begin{tabular}{lccc}
\hline
\hline
Configuration                  &Default             &1$^{st}$ Study\cite{LL:JMF-005}           &2$^{nd}$ Study\cite{LL:GAIA.ASF.TCN.PLM.00884}\\
\hline
VPU Parameter                  &&&\\
\hline
 EXT\_IN\_BUFFER [pixels]      &6                   &6                                       &6  \\
 EXT\_WH         [extremities] &25                  &128                                     &127 \\
 DELTA\_MAG      [g\_mag]      &128                 &32                                      &90  \\
 MAG\_AL\_LUT    [g\_mag]      &default PSF model   &default PSF model                       &PSF, quilting effect included \\
 MAG\_AC\_LUT    [g\_mag]      &default PSF model   &default PSF model                       &PSF, quilting effect included  \\
\hline
\end{tabular}
\end{center}
\end{table}

\begin{figure}[h!]
\begin{center}
\includegraphics[width=0.7\hsize]{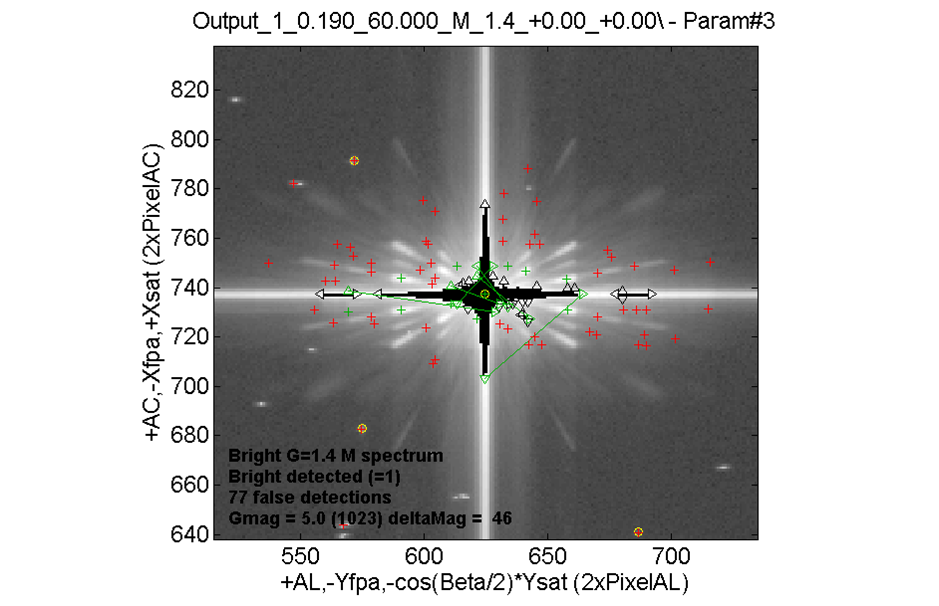}
\end{center}
\caption{Example of simulated image used as input for the 2$^{nd}$ Study (Image courtesy of ADS\cite{LL:GAIA.ASF.TCN.PLM.00884}) corresponding to a star of magnitude $G=1.4$ and M spectral type. The PSF model includes the quilting effect. Overimposed to the input image the result from the VPU simulations is plotted. The yellow circles correspond to true stars in the input image, the black triangles to detected extremities, while the green triangles are extremities that have been actually matched. The green crosses correspond to observed objects resulting from the extremities matching process, while the red crosses indicate a faint-object observed. The horizontal axis represents {samples the AL direction and the vertical axis is in AC samples.}
\label{fig:psfQuilting}}
\end{figure}

\section{VPU parameter optimisation}
\label{sec:vpu}
In 2012, we conducted the first study to explore the limits of the VPU saturated star algorithms\cite{LL:JMF-005}. We made extensive use of the VPU Algorithm prototype version 2.5 developed by ADS to carry out comprehensive on-ground simulations to determine optimised values of those five parameters to extend the bright-star detection limit. The SM and AF input images required were simulated using a PSF simulator developed at the Gaia Science Operations Centre (SOC), the "VPU Image Generator" tool\cite{LL:JMF-003}. The PSF model\cite{LL:AMF-002} used to generate the stars includes Gaussian read-out noise and Poisson shot-noise. A set of seventeen stars was created with half-magnitude steps in the range $0.0 \le G \le 8.0$, plus one star at the default magnitude limit, $G=5.7$. The stars were simulated as solar-type G2V stars in FOV1, located at the same pixel position, without AC drift due to spin axis precession, and only one noise configuration was generated for each star in this first study. The stars in the range $0.0 \le G$ \textless  $5.7$, outside the nominal magnitude range of Gaia, were the stars of interest. While the stars in the range from $5.7 \le G \le 8.0$ were used as a control sample to check whether any side effect was introduced in the detection of stars in the nominal magnitude range.

The results of this study was an optimised set of VPU parameters, which is shown in the second column of Table~\ref{tab:Param}. We found that EXT\_WH is the most relevant parameter to establish the detection limit. The default value of EXT\_IN\_BUFFER was found to be adequate. DELTA\_MAG acts as a filter to avoid multiple observations of the same bright star, i.e. when more than one observation (window class\_0) is assigned. We did not explore changes to "MAG\_AL\_LUT" and "MAG\_AC\_LUT", because the maximum distance between 2 matching extremities used as input for the LUTs is 32 pixels, that corresponds approximately to a magnitude $G\sim4$.

\begin{figure}[h!]
\begin{center}
\includegraphics[width=0.023\hsize,height=0.14\vsize]{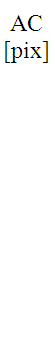}
\includegraphics[width=0.48\hsize,height=0.2\vsize]{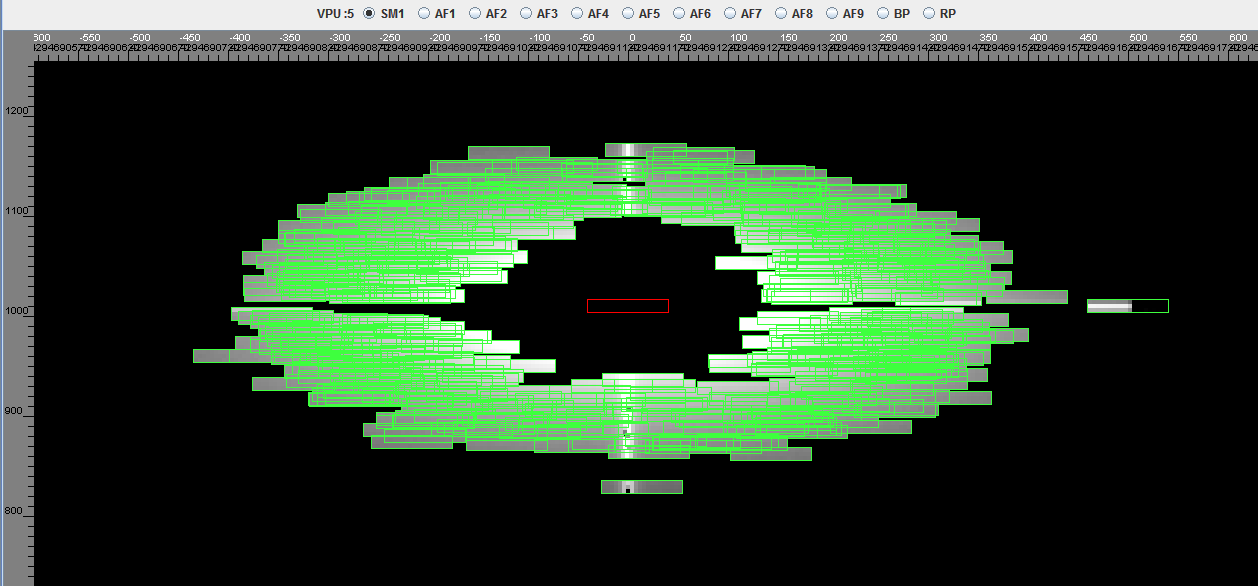}
\includegraphics[width=0.48\hsize,height=0.204\vsize]{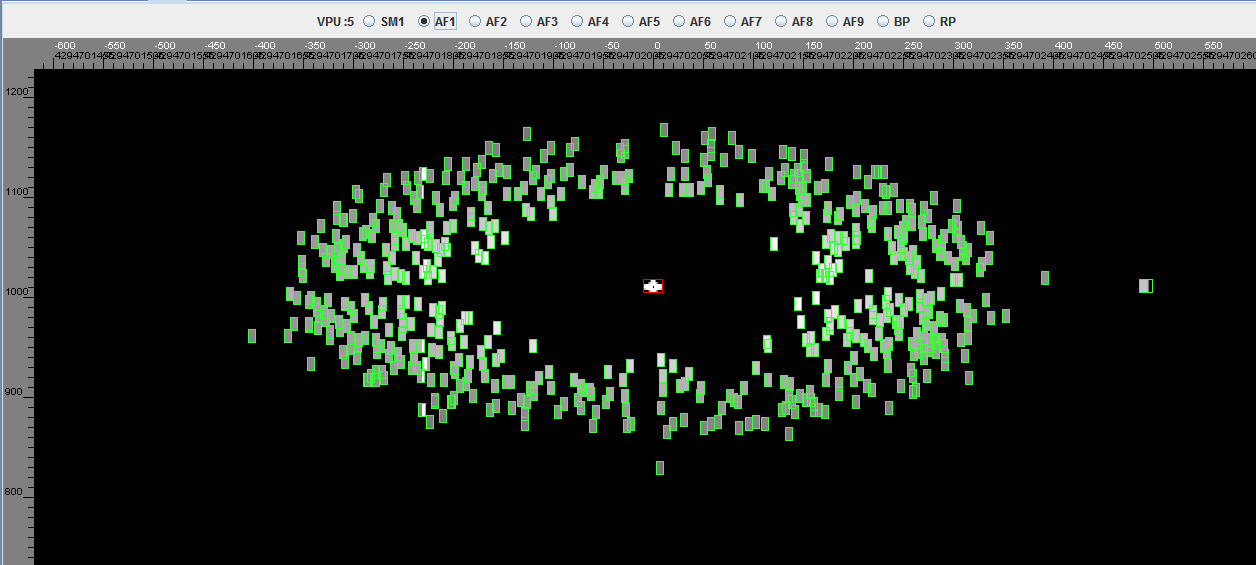}
\hspace{0.3\hsize}
\includegraphics[width=0.075\hsize]{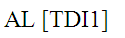}
\end{center}
\caption{G2V type star of magnitude $G = 1.5$ detected on the VPA SW Prototype simulations using the best configuration of the first study (Table~\ref{tab:Param}). SM image is shown on the left and AF1 image on right. The SM S/W processing algorithm detects one bright star from the extremities. The observation window is properly assigned to the bright object and tracked through all the FPA (red squared window in the center of the cloud). The star is surrounded by a cloud "false-object" observations associated to photon shot noise in the star PSF wings. A charge injection feature can be seen in some of the "false-object" windows in AF1. 
\label{fig:g2V_G1_5_DeltaMag}}
\end{figure}

With this VPU configuration, the bright magnitude limit was $G\approx2$, hence very promising. An example of one simulated observation using this configuration is shown in Figure~\ref{fig:g2V_G1_5_DeltaMag}. It can be noticed that the observation of a bright star is surrounded by a cloud of "false-object" observations, which are caused by photon shot noise in the PSF wings.

A second study\cite{LL:GAIA.ASF.TCN.PLM.00884} was carried out by ADS in coordination with the SOC to verify the results and its implementation in the Avionics Verification Module (AVM) in a real VPU unit with SW version 2.6. {This study considered additional aspects that may affect} the bright-star detection limit. Among these were (1) the star spectral type (B1V, G2V and M6V) (2) the star centroid subpixel location (0.25 pix/step) (3) star transits with the maximum AC speed (190 mas/s) and AL speed (60002 mas/s) (4) a more realistic PSF model that includes the quilting effect (Figure~\ref{fig:psfQuilting}) (5) a background field of faint stars and cosmic rays in the simulated images and (6) {two different Wave-Front Errors (WFE) cases, correspondent to row 1 and 4, both on FOV1}. Several configurations of the VPU parameters were exercised and the MAG\_AL\_LUT and MAG\_AC\_LUT tables were calibrated for each VPU row using the Code V WFE estimates and the PSF simulations.
 
The results of the second study essentially confirmed the results of the first study, refined the VPU parameters configuration (third column in Table~\ref{tab:Param}), and allowed to estimate the error on the magnitude limit found. The expected bright magnitude limit in this case was $G = 2.0 \pm 0.6$. 

This second study also confirmed the presence of a cloud of "false-object" observations around the bright star (red crosses not linked to any real star, yellow circle on Figure \ref{fig:psfQuilting}). It also confirmed that in many cases several extremities crossmatches survived for the same star, which implies more than one observation {(window class\_0)} assigned,{ leading} to several observations of the same star.  

Finally, the implementability of the parameters on a real VPU unit was demonstrated without any relevant side effect. The VPU parameter configuration resulting from this study was adopted as the baseline for the commissioning phase.

\section{SIF Imaging}\label{sec:sif}
\subsection{Motivation}
As indicated by our simulations, Gaia {would be} unable to automatically observe stars brighter than $G\sim2$ even with the optimised set of VPU parameters. The number of stars with $G < 2$ is small\cite{1997ESASP1200.....P} (approximately 60) and when one of those transits the FPA, its information is read in the SM and stored in the SM Raw Data Buffer, just as for any other star. However, for those very bright stars, the extension of the saturated area exceeds the maximum distance allowed for the cross-matching process. Therefore, no window is assigned to the star and it is not observed in AF. Only the cloud of "false-detections" around the star will be observed.

Using those spurious detections around the brightest star detected ($G = 1.5$) in our simulation\cite{LL:JMF-005}, we did an approximate reconstruction of its SM image. Applying a custom version of the Cram\'er-Rao centroiding precision estimator, we estimated that the astrometric information in a single SM image would be $\sim20.8~\mu as$. According to the simulations done using the Gaia Accuracy Analysis Tool (GAAT)\cite{LL:JdB-053}, this precision is equivalent to that obtained for a $G \sim 11.5$ star, after combination of the information contained in all the AF1-9 windows obtained in a single transit. As a consequence, SM images can contain sufficient astrometric information to allow the extension of the Gaia bright limit for astrometry, provided that systematic errors of the centroiding for these images can be mitigated. This motivated us to pursue a technique that allow us to recover the information in the SM Raw Data Buffers, the "SIF imaging" technique.

\subsection{Principle}
The Service Interface Function (SIF)\cite{LL:GAIA.ASU.ICD.PLM.00012} provides access to the SDRAM memory contents in the Processor Board of the VPU. Originally the SIF was designed as a pure engineering tool, which allows obtaining detailed information about the VPU processes. This information can be used for a wide variety of purposes, {e.g. }SW debugging or extracting data not available in the nominal TM ({for instance} post scan pixels or braking samples). {We} used the SIF to read the FVP stored in the SM Raw Data Buffer and the AF1 information on the SVP Raw Data Buffer. The FVPs were read either from SM1 or SM2 data buffer depending on the FOV the object transited through. By reconstructing the SM image from the samples read with the SIF, we get a full image {of the} SM CCD for the duration the SIF was sampling the buffer. The data gathered from the SVP Raw Data Buffer for AF1 allows us to verify when a successful cross-match (i.e. an on-board confirmation) has happened on-board. 

In order to avoid generating unmanageable amounts of telemetry, the SIF imaging technique requires to predict the transit time of the object of interest in the focal plane. The SOC designed the Gaia Observing Schedule Tool (GOST)\cite{LL:WOM-068}. During commissioning, a customised version of GOST allowed us to predict the object transit times in both FoVs with an accuracy of $\sim0.2$ s and an error in AC position of $\sim10$ AC pixels\cite{LL:JSA-002} . 

\subsection{Commissioning results}
We used the SIF imaging technique to determine the actual bright limit for source observation onboard Gaia with the optimised VPU parameters. We predicted the transit times and AC locations of very bright stars ($G<5.7$) and commanded SIF imaging of one VPU for several seconds around the expected transit time.
The data obtained from the SIF was reconstructed to obtain the images of all the targeted objects and to determine the actual time and AC position of the star image as seen by Gaia.  By inspecting the nominal data products of Gaia (SP1) in the vicinity of the very bright star event, we could determine whether the target object was observed by the on-board algorithms. This yields an assessment of the Gaia observation efficiency in the $G<5.7$ magnitude range.

\begin{figure}[h!]
  \begin{center}
    \includegraphics[width=0.9\hsize]{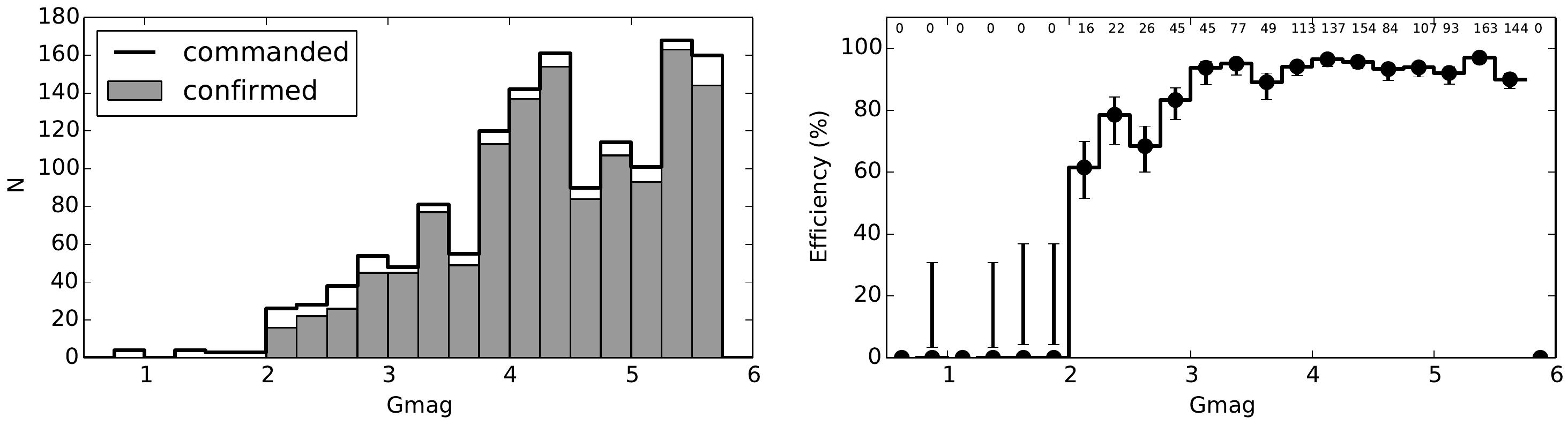}
    \hspace{0.35\hsize}
  \end{center}
  \caption{On-board confirmation rate as obtained using the SIF imaging technique during the commissioning phase for objects brighter than $G < 5.7$. Left: Magnitude histogram of commanded and confirmed stars. Right: Onboard source confirmation efficiency as a function of $G$ magnitude. Black bars show the uncertainties corresponding to binomial statistics. The number in the upper part indicates the number of confirmed sources in the respective bin. The faintest bin ($G$ = 5.75 - 6.0) is empty because no observation was commanded there.}
\label{fig:efficiency}
\end{figure}

The data gathered using SIF targeted observations in SM and AF1 during the Gaia commissioning phase comprises 1400 very bright star transits. Of these, 1275 were confirmed by the on-board algorithms and consequently observed by Gaia. Figure~\ref{fig:efficiency} shows the on-board observation rate as a function of the star $G$ magnitude. In the $G$ = [3.0 - 5.7] range, 1166 stars out of 1240 stars have been observed on-board, i.e. the corresponding SP1 packet was generated, which corresponds to a rate of $94.0\%^{+0.6\%}_{-0.7\%}$\footnote{The quoted uncertainties correspond to 68 \% confidence intervals and were obtained under the assumption of binomial statistics}. The efficiency drops rapidly for brighter stars and in the $G$ = [2.0 - 3.0] range it is $74.7\%^{+3.3\%}_{-3.9\%}$. Only 14 stars brighter than $G$ = 2 were observed with SIF and none was confirmed by the on-board algorithms, thus the efficiency here is smaller than $12\%$.

\begin{figure}
  \begin{center}
    \includegraphics[width=0.8\hsize]{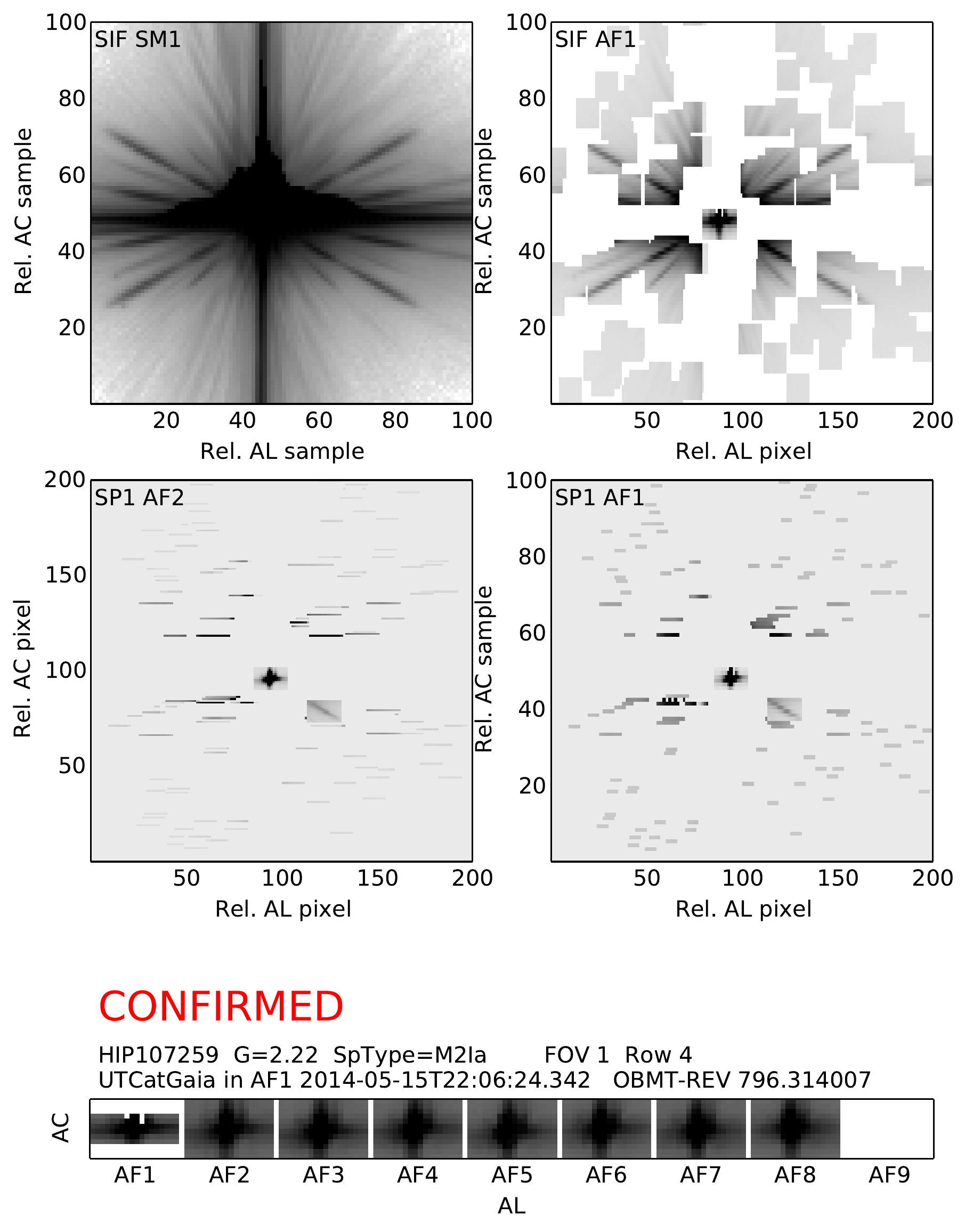}
    \hspace{0.35\hsize}
  \end{center}
  \caption{Plot showing the data gathered during the commissioning for an object of G=2.22 that transited the FPA row 4 FOV 1 observed both by the autonomous observation algorithms using the VPU configuration parameters from Table\_\ref{tab:Param} and using the SIF imaging technique. Top row: data gathered using the SIF imaging technique. Left: SM image reconstructed from the FVP data. Right: image reconstructed from the AF1 SVPs on the same period of time. Middle row: image reconstructed from the actual observations made by Gaia on AF2 (left) and on AF1 (right) centered on the bright object. Bottom row: Bright object data, comprising its identifier, spectral type, FOV and row it transited, below it, its transit time. Below the transit time, the observations made by Gaia of the bright object in all the AF CCDs are shown. (Notice the absence of data in AF9, because in row 4 it has been replaced by a WFS).}
\label{fig:efficiency2}
\end{figure}

\section{VO Synchronised Observations}
\label{sec:vo}
\subsection{Motivation}\label{sec:VOmot}
The SIF imaging technique can record images from the SM CCDs for any star, particularly for the ones that are too bright to be detected by the VPU algorithms. However, when those extremely bright objects transit the FPA {no other CCD is acquiring} data of them. Even though core saturation is strong for such stars, {simulations showed that} the astrometric information in the wings is comparable to regular bright stars $G = [6.0, 13.0]$. BP/RP low resolution spectra are also saturated in the core columns, but plenty of  {signal is present at a few samples distance }in AC direction. High precision astrometry and absolute photometry could be possible for $G < 2.0$ stars, provided that the VPU {can} assign windows and resources to observe them. This {was} the motivation to {devise} a means of recording such data, the "VO Synchronised Observations" (VO-Sync) technique.

\subsection{Principle}\label{sec:VOprinc}
The VPU algorithms include a SW tool designed to introduce, in parallel to SM1 and SM2 detected objects, an additional set of externally defined objects that can be inserted in the buffer of "real" objects to be submitted to the rest of the processing chain for observation in the FPA. These are the called Virtual Objects (VO) {and} are intended to force the observation of windows placed at known locations on AF1. The VOs are observed and processed as any other object on the FPA. Once {acquired}, the same rules are applied to them as to any other "real" object {and the data} are provided by the VPU in {the form of} star packets. VOs are defined by the structured parameter VO\_DEFINITION of the VPU. This parameter defines the VO observation time, AC coordinate, magnitude, observation priority, and class of sampling, among others.

\subsection{Commissioning results}\label{sec:VOcom}
The idea behind the VO-Sync technique is to predict a bright object transit and to place a class\_0 VO window on top of it. However, that was never the original purpose of VOs. They were developed as a debugging tool capable of testing complex object scenarios and a way to regularly sample the background by randomly placing observing windows in the sky, according to a number of predefined VO patterns. The test done during commissioning was to insert a series of 3x2 (ALxAC) arrays of class\_0 VOs in the routine VO pattern, such that each {array falls} on top of a bright star.

This observation strategy requires, apart from a careful tuning of the VO pattern, transit predictions accurate to $\pm6$ pixels ({initially} $\pm12$ pix) in AC and to $\pm9$ pixels ({initially} $\pm27$ pix) in AL. That is, predictions accurate to 6-27 ms (AL) and 1-2 arcsec (AC) are required. {This is only} possible if the Attitude and Orbit Control System (AOCS) jitter is smaller than these {values},{ which themselves are }much smaller than the requirements. In addition, it requires an accurate modeling of the AOCS systematic errors,{ e.g., misalignment between star tracker and FPA, differential distortion, and precession effects. }We showed this is possible by modeling the GOST prediction errors measured over 2 days with sinusoidal terms\cite{LL:JSA-004_JSA}. These terms were used to predict the transits of stars with a time horizon up to one week. The RMS of the resulting prediction error is $\sim5$ pixels in AL and $\sim8$ pixels in AC. This is sufficient to catch the majority of very bright star (VBS) events with VOs. We expect to minimise these prediction errors even further by improving our model, e.g. by using stellar photocentres and {dependencies on sky position and scan law.}

\begin{figure}[!ht]
  \begin{center}
		\includegraphics[width=0.023\hsize,height=0.1\vsize]{fig2.png}
    \includegraphics[width=0.40\hsize]{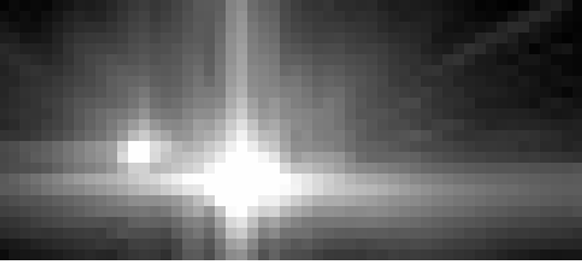}
    \hspace{0.80\hsize}
		\includegraphics[width=0.023\hsize,height=0.1\vsize]{fig2.png}
    \includegraphics[width=0.80\hsize]{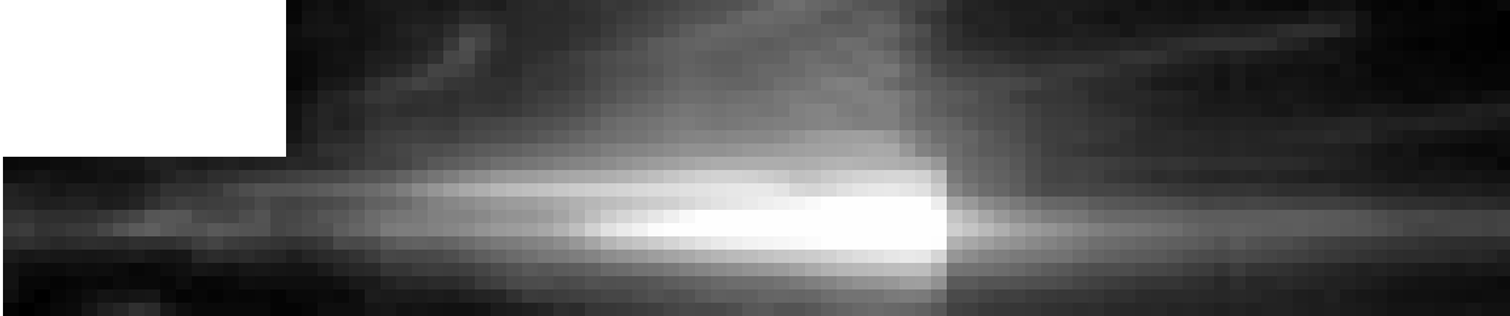}
    \hspace{0.80\hsize}
   	\includegraphics[width=0.023\hsize,height=0.1\vsize]{fig2.png}
    \includegraphics[width=0.80\hsize]{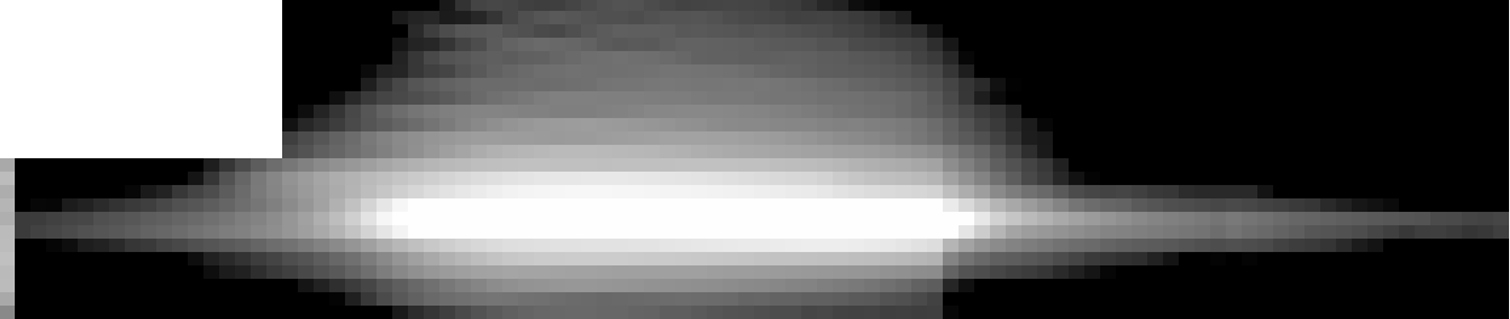}
    \hspace{0.80\hsize}
    \includegraphics[width=0.08\hsize,height=0.01\vsize]{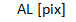}
  \end{center}
  \caption{Images of HIP54061 ($G=1.53$) captured using the VO-Sync observation technique. The AF2 image is the image located at the top row, in the middle row is the BP image, and in the bottom is the RP image. {See \url{http://www.cosmos.esa.int/web/gaia/iow\_20140605} for nominal examples of BP and RP images}.}
\label{fig:VO1.5}
\end{figure}

Figure \ref{fig:VO1.5} shows the images captured during the commissioning phase of HIP54061 (alf UMa, $G=1.5$) using the VO-Sync technique. Further tests and refinements of this technique are needed before it can be considered during regular operations.

\section{Further improvements}
\label{sec:further}
The {achievable} bright magnitude limit is mainly driven by the configuration of the extremities cross-matching process, {in particular by the parameter EXT\_WH, whose optimised value is already in use}. Therefore, further extending the bright magnitude limit {in this way }seems not to be feasible. However, {by} calibrating the other parameters involved in the cross-match {it is} possible to improve the performances of the algorithms, i.e. improve the magnitude estimation for bright stars and diminish the number of multiple detections.

The saturated area in {SM images of very bright stars show} an asymmetry in the AC direction (Figure \ref{fig:efficiency2}, top row{ left}). The default MAG\_AL\_LUT and MAG\_AC\_LUT parameters were computed assuming a symmetric PSF, hence a symmetric saturated area. We envisage that including this asymmetry in the definition of those parameters would improve the precision on the magnitude estimation of the extremities. In principle, this improvement would only be evident for stars with $G~\textgreater~4$ (see Sect.\ref{sec:vpu}). 

Furthermore, a smaller spread in the magnitude estimation of the extremities would allow to perform a better calibration of DELTA\_MAG. {A} better selection of cross-matches coming from the same star would then be possible. Therefore, in that case we expect a slight decrease on the number of multiple observations coming from one star. The observation of a bright star window (class\_0) consumes half of the observation resources available in AF {for the extension of the window (18 pixels)}. Therefore, avoiding multiple observations of a single bright star would increase the resources available to observe faint objects simultaneously. 

A third study is planned {to run the} VPU prototype {with} real SM and AF1 images, obtained using the SIF imaging, as input. The PSF {varies} from row to row, {hence}, this study should be done for each VPU independently. I{f we find that there is room for significant} improvement, {we will }derive a new set of VPU parameters that {allows us} to {optimise} the algorithm performance for individual VPUs.

\section{Conclusions}\label{sec:conclusion}
The performance using the VPU parameter configuration resulting from the simulations\cite{LL:JMF-005, LL:GAIA.ASF.TCN.PLM.00884} , intended to extend the Gaia bright limit, were baselined and tested during the commissioning phase. It was demonstrated that this configuration allows Gaia to observe stars in the $G = 3.0~$-$~5.7$ range with an efficiency close to 94\% and stars in the $G = 2.0~$-$~3.0$ range with an efficiency close to 75\% \cite{LL:JMF-006, LL:JSA-003}. The astrometric performance achievable with these data remains to be quantified, which is work in progress.

Furthermore, two targeted observations techniques were designed and tested in the commissioning phase. The SIF imaging technique was extensively used, first to recover the data from AF1 for objects actually observed by Gaia, and second to recover an image from SM for any star. The data quality for those stars with $G<5.7$, which suffer from heavy saturation both in SM and AF, remains to be quantified. However, according to our simulations the astrometric information contained in a single SM image of such a very bright object could be $\sim20.8~\mu as$ . The number of stars with $G < 3$ is only 230, thus, targeting SIF imaging observations for those bright objects during the mission would allow Gaia astrometry to{ become complete at the bright end}. This is a planned activity.

Finally, the VO synchronised technique was tested and its feasibility has been proven. It has demonstrated to be a powerful observation tool to retrieve a full observation (astrometric, photometric and spectroscopic data) for any object, even for those not observed through the autonomous observation algorithms. This technique requires to minimise the transit prediction errors {and} further testing is needed before{ it can be} used during routine operations.

Gaia entered in nominal mission operations {in July 2014} and the VPU parameter configuration presented here was {adopted} for the nominal mission. Consequently,{ the magnitude range of Gaia was extended }to $G=[3.0-20.0]$. Stars with $G<3$ will systematically be observed with the {SIF imaging technique}\footnote{Additional information can be found at \url{http://www.cosmos.esa.int/web/gaia/science-performance}.}.

\acknowledgments
The authors wish to thank the Gaia Science Team and Data Processing and Analysis Consortium (DPAC) for their support and the Airbus Defence \& Space (ADS) for their collaboration and their assessment. All three are gratefully acknowledged for their contribution and access to their documentation. Some concepts and ideas presented here come from those sources. And part of the supporting material used in this work has been provided by them.
 
We also kindly acknowledge the Gaia DPCE and the MOC operations teams for their continuous support and good will to perform the 1400 additional observations during the commissioning phase.

\end{document}